\documentclass[12pt]{article}
\usepackage{amssymb}
\usepackage{amsfonts}
\usepackage{amsmath,amsthm}
\usepackage{graphics}
\oddsidemargin 10mm \numberwithin{equation}{section}
\def\be{\begin{equation}}
\def\ee{\end{equation}}
\begin{document}
\begin{center}
{{\bf {Spherically symmetric
 Jordan-Brans-Dicke quantum gravity with de Broglie Bohm pilot wave perspective}}\\
 \vskip 0.50 cm
  {{Hossein Ghaffarnejad}}\footnote{E-mail address:
hghafarnejad@ yahoo.com. }\vskip 0.1 cm \textit{  Department of
Physics, Semnan University, P.O.Box 35195-363, Iran}}
\end{center}
\vspace{0.1cm} \begin{abstract} We obtain two dimensional analogue
of the Jordan-Brans-Dicke (JBD) gravity action described in four
dimensional spherically symmetric curved space time metric. There
will be two scalar fields, namely, the Brans Dicke (BD) $\phi$ and
scale factor of 2-sphere part of the space time $\psi.$ We obtained
suitable duality transformations between $(\psi,\phi)$ and
$(\rho,S)$ where $\rho$ and $S$ is respectively amplitude and phase
part of the corresponding de Broglie pilot wave function
$\Psi(\rho,S)=\sqrt{\rho}e^{iS}.$ Covariant conservation of
mass-energy current density of particles ensemble
$J_a=\rho\partial_aS,$ is established by applying a particular
dynamical conformal frame described by $(\rho,S).$
 \end{abstract}
 \section{Introduction}
Over half a century of collective study has not diminished the
fascination of searching for a consistent theory of quantum
gravity [1].\\
The perturbative quantum field theory in curved space time, [2,3,4]
schemes foundered on intractable ultraviolet divergences and gave
way to super-gravity, the super-symmetric extension of standard
general relativity.  In spite of initial optimism, this approach
succumbed to the same disease and was eventually replaced by the far
more ambitious superstring theories [5]. Superstring theory is now
the dominant quantum gravity programme in terms of the number of
personnel involved and
the number of published papers, per year, per unit researcher.\\
The nonperturbative canonical quantum gravity [6] or quantum
geometry attempts for quantizing the metric variables where rather
nave and took on various forms according to how the intrinsic
constraints of classical general relativity are handled. In the most
popular approach, the constraints are imposed on the state vectors
and give rise to the famous Wheeler-De Witt equation arguably one of
the most elegant equations in theoretical physics, and certainly one
of the
most mathematically ill-defined.\\
The enormous difficulty of the canonical quantum gravity scheme
eventually caused it to go into something of a decline, until new
life was imparted with Ashtekar`s discovery of a set of variables in
which the constraint equations simplify significantly. This scheme
slowly morphed into ``loop quantum gravity``: an approach which has,
for the first time, allowed real insight into what a nonperturbative
quantization of general relativity might look like. A number of
genuine results were obtained, but it became slowly apparent that
the old problems with the Wheeler De Witt equation were still there
in transmuted form and the critical Hamiltonian constraint was still
ill-defined. In summary, the canonical quantum gravity and
particularly loop quantum gravity has self-contained treaties and it
may to be hopeful (see [7] and references
therein).\\
Quantum general relativity or canonical quantum gravity is based to
two fundamental building blocks of modern physics, namely, general
covariance (the general relativistic principle of background
independence) and the uncertainty principle of Copenhagen quantum
mechanics. However the well known Copenhagen quantum mechanics has
still many fundamental problems: non-locality, non-causality, the
measurement problem,  the nature of reality and etc. The
Bohr-Einstein debate on these problems is followed by other
researchers and whose challenge reduces to the de Broglie-Bohm
casual quantum mechanics [8-14]. The de Brogle-Bohm ontology of the
quantum physics presents new features of quantum gravity theory (see
for instance [15-18]). In the standard form of this theory, the
classical gravity should be viewed as a classical field containing
 quantum trajectories originated from
quantum potential.\\
The author exercised conformaly flat space time version of the JBD
gravity [19] previously and obtained whose de Broglie-Bohm particle
interpretation in [20]. In the present paper, we use JBD scalar
tensor gravity described by spherically symmetric curved space-times
and obtain whose corresponding de Borglie-Bohm quantum gravity in
which amplitude and phase of the pilot wave can be described in
terms of the BD scalar field and scale factor of 2-sphere part of
the metric. As a future work, results of the paper can be  used to
study physical features of spherically symmetric dynamical space
times such as black holes evaporation, thermodynamics and etc.
 \section{Spherically symmetric JBD space time}
Let us start with JBD gravity theory [19] given by \be
I_{JBD}=\frac{1}{16\pi}\int d^4x\sqrt{g}\{\phi
R-\frac{\omega}{\phi}g^{\mu\nu}\partial_{\mu}\phi\partial_{\nu}\phi
 \}.\ee Using a dimensionless scalar field as
  \be\sigma=(2\omega+3)^{\frac{1}{2}}\ln{G\phi},\ee  and particular conformal metric  transformation \be\bar{g}_{\mu\nu}
 = g_{\mu\nu}\exp{\left(\frac{\sigma}{\sqrt{2\omega+3}}\right)}=
 g_{\mu\nu}G\phi\ee  the action (2.1) leads to the following form.
 \be \bar{I}_{JBD}=\frac{1}{16\pi G}\int \sqrt{\bar{g}}d^4x\{\bar{R}-\frac{1}{2}
 \bar{g}^{\mu\nu}\partial_{\mu}\sigma\partial_{\nu}\sigma
 \}\ee which describes minimally coupled scalar-tensor gravity theory where $\bar{I}_{EH}=\frac{1}{16\pi G}\int
 \sqrt{\bar{g}}d^4x\bar{R},$ is well known Einstein-Hilbert action
 functional. Usually $\bar{g}_{\mu\nu}$ and $g_{\mu\nu}$ is called `Einstein` and
`Jordan` frames respectively [21]. $G$ is named the Newton`s
gravitational coupling constant.  Also $g$ $(\bar{g})$ is absolute
value of determinant of the metric $g_{\mu\nu}$ $(\bar{g}_{\mu\nu})$
with Lorentzian signature $(-,+,+,+).$ $\omega$ is dimensionless BD
parameter and whose present limits based on
 time-delay experiments [22-25] require
$\omega\geq4\times10^4.$  When $\omega\to\infty$, then the JBD
gravity theory leads to the Einstein`s general relativity theory and
$\omega=-1$ is due to a fundamental symmetry of strings [26]. This
is a symmetry of string amplitudes which relates large and small
radius of compactification. Also negative values of the BD parameter
$\omega$ come from Kauza-Klein theory, when these alternative
theories in (4+h) dimensions reduce to a generalized JBD theory
after the dimensional reduction in the zero modes approximation
such as $\omega=-(1+\frac{1}{h})$ [27].\\
General form of spherically symmetric curved space time metric is
given by
 \begin{equation}
d s^2=g_{a b}dx^{a}dx^{b}+\psi^{2} \left(d\theta^{2}+\sin^{2}\theta
d\varphi^{2}\right),\label{1.2}
\end{equation}
where 2-sphere conformal factor $\psi^2$ is called as dilaton field
and $g_{ab}$ is described in terms of two dimensional coordinates
$x^{a}$ with $a \equiv 1,2$. Thomi et al [28] obtained previously
two dimensional analog of the Einstein-Hilbert action by using (2.5)
such as follows. \be \bar{I}_{EH}=\frac{1}{16\pi G}\int
d^4x\sqrt{^{[4]}\bar{g}}^{[4]}\bar{R}=\frac{1}{2G}\int
d^2x\sqrt{\bar{g}}\{1+{\bar{g}}^{ab}\partial_a\bar{\psi}\partial_b\bar{\psi}+
\frac{1}{2} \bar{R}{\bar{\psi}}^2\}\ee where `over-bar` denotes to
the Einstein frame. $^{[4]}g$ ($g$) and $^{[4]}R$ ($R$) is absolute
value of determinant of the four (two) dimensional metric and Ricci
scalar in four (two) dimensional space time respectively. Applying
(2.5) and (2.6), one can obtain two dimensional analogue of the
action (2.4) described in the Einstein frame as \be
\bar{I}_{JBD}=\frac{1}{2G}\int
d^2x\sqrt{\bar{g}}\{1+{\bar{g}}^{ab}\partial_a\bar{\psi}\partial_b\bar{\psi}+
\frac{1}{2}
\bar{R}{\bar{\psi}}^2-\frac{\bar{\psi}^2}{4}\bar{g}^{ab}\partial_a\bar\sigma\partial_b\bar\sigma\}.\ee
Using the conformal transformation (2.3), we will have \be
\sqrt{\bar
g}=\sqrt{g}\exp\left(\frac{\sigma}{\sqrt{2\omega+3}}\right),~~~\bar{g}^{ab}=g^{ab}\exp\left(\frac{-\sigma}{\sqrt{2\omega+3}}\right)\ee
and \be
\bar{\psi}=\psi\exp\left(\frac{\sigma}{2\sqrt{2\omega+3}}\right),~
~~\bar\sigma=\sigma,\ee with
\be\bar{R}=\exp\left(-\frac{\sigma}{\sqrt{2\omega+3}}\right)\left[R-\frac{g^{ab}\partial_a\partial_b\sigma}{\sqrt{2\omega+3}}\right].\ee
Applying (2.2), (2.8), (2.9) and (2.10), the action (2.7) is given
exactly in the Jordan frame such as follows. \be
I_{JBD}^{Jordan}[\phi,\psi,g_{ab}]=\frac{1}{2}\int
d^2x\sqrt{g}\{\phi+\frac{1}{2}\phi\psi^2 R+\phi
g^{ab}\partial_a\psi\partial_b\psi$$$$+2\psi
g^{ab}\partial_a\psi\partial_b\phi-\frac{\omega\psi^2}{2\phi}g^{ab}\partial_a\phi\partial_b\phi\}\ee
in which $R$ is Ricci scalar of the metric $g_{ab}$ defined in
Jordan frame, and also we eliminated divergence-less terms. Varying
the action (2.11), with respect to the fields $g^{ab},$ $\psi$ and
$\phi$ the corresponding field equations are obtained respectively
as \be
T_{ab}=\phi\partial_a\psi\partial_b\psi+2\psi\partial_a\psi\partial_b\phi-\frac{\omega\psi^2}{2\phi}\partial_a\phi\partial_b\phi-\partial_a
[\sqrt{g}\partial_b(\phi\psi^2)]/2\sqrt{g}$$$$-\frac{1}{2}g_{ab}\{\phi+
g^{cd}[\phi\partial_c\psi\partial_d\psi+2\psi\partial_c\psi\partial_d\phi-\omega\psi^2\partial_c\phi\partial_d\phi/2\phi-\Box(\phi\psi^2)]\},\ee
\be \frac{1}{2}\phi\psi
R-g^{cd}[\partial_c\phi\partial_d\psi+\omega\psi
\partial_c\phi\partial_d\phi/2\phi]-\psi\Box\phi-\phi\Box\psi=0\ee
and \be
1+\frac{1}{2}\psi^2R+g^{cd}[\partial_c\psi\partial_d\psi-\omega\psi^2
\partial_c\phi\partial_d\phi/2\phi^2+2\omega\psi
\partial_c\psi\partial_d\phi/\phi]+\omega\psi^2\Box\phi/\phi-\Box\psi^2=0\ee
where we defined $\Box=\partial_a(\sqrt{g}g^{ab}\partial_b)/\sqrt{g}.$\\
Using the transformations \be
\phi\psi^2=\alpha\rho,~~~\psi=l_p\rho^{\beta}e^{\gamma S}\ee with
\be
\alpha=\frac{3+2\omega}{4(2+\omega)},~~~\beta=\frac{1+\omega}{3+2\omega},~~~~\gamma^2=-\frac{2(2+\omega)}{(3+2\omega)^2}\ee
 the action (2.11) become \be
I^{Jordan}_{JBD}[\rho,S,g_{ab}]=\frac{1}{2}\int d^2x\sqrt{g}\{\rho
g^{ab}\partial_aS\partial_bS+\frac{g^{ab}\partial_a\rho\partial_b\rho}{4\rho}+\alpha\rho
R+\frac{2\alpha\rho^{1-2\beta}}{l_p^2e^{2\gamma S}}\}\ee where
$l_p=(16\pi G)^{1/2}$ with $c=\hbar=1,$ is the Planck length, and
$(\rho,S)$ are dimensionless real scalar fields. Varying the action
(2.17) with respect to the field $\rho,$ one can obtain
Hamilton-Jacobi equation as \be g^{ab}\partial_a
S\partial_bS=-\alpha R+\frac{(2\beta-1)e^{-2\gamma
S}}{\rho^{2\beta}l_p^2}+\frac{\Box\sqrt{\rho}}{\sqrt{\rho}}.\ee
Varying the action (2.17) with respect to the field $S,$ we obtain
covariant divergence of the mass-energy current density of the
particles ensemble $J_a=\rho\partial_aS$ such that    \be
\frac{1}{\sqrt{g}}\partial_a\{\rho\sqrt{g}g^{ab}\partial_bS\}=
-\gamma\rho^{1-2\beta} e^{-2\gamma S}/l_p^2.\ee The above equation
shows covariant conservation of the mass-energy current density
$J_a$ only at the particular Einstein`s regime $\gamma=0,$ namely
$\omega\to+\infty.$ In other words nonzero value of RHS of the above
equation presents particle creation coming from Bohm quantum
potential. However, with arbitrary nonzero value of $\alpha$
$(\omega\nrightarrow\infty)$, the covariant conservation of this
current density can be still established  by applying conformal
frame  $\tilde{g}_{ab}=\Omega^2g_{ab}$ where \be
\tilde{\Box}\ln\Omega=-\rho^{-2\beta}e^{-2\gamma S}/l_p^2.\ee
       In the latter frame the action (2.17) is transformed to \be
      \tilde{I}^{Jordan}_{JBD}[\rho,S,\tilde{g}_{ab}]=\frac{1}{2}\int d^2x\sqrt{\tilde{g}}\{\rho
\tilde{g}^{ab}\partial_aS\partial_bS+\frac{\tilde{g}^{ab}\partial_a\rho\partial_b\rho}{4\rho}+\alpha\rho
\tilde{R}\}\ee in which the corresponding Hamilton-Jacobi equation
and the covariant conservation equation are obtained as \be
\tilde{g}^{ab}\partial_aS\partial_bS=-\alpha
\tilde{R}+\frac{\tilde{\Box}\sqrt{\rho}}{\sqrt{\rho}}\ee and \be
\frac{1}{\sqrt{\tilde{g}}}\partial_a\{\rho\sqrt{\tilde{g}}\tilde{g}^{ab}\partial_bS\}=0.\ee
Defining de Borglie-Bohm pilot wave of the gravitational system as
$\Psi=\sqrt{\rho}e^{iS},$  the action (2.21) can be rewritten as \be
\tilde{I}^{Jordan}_{JBD}[\Psi,\tilde{g}_{ab}]=\frac{1}{2}\int
d^2x\sqrt{\tilde{g}}\{\tilde{g}^{ab}\partial_a\Psi\partial_b\Psi^{*}+\alpha\Psi\Psi^*\tilde{R}\}\ee
where $\Psi^*=\sqrt{\rho}e^{-iS}$ is complex conjugate of the scalar
field $\Psi.$ In order to obtain the WKB approximation, in the
classical limits where there is a wave packet of width much greater
than the wave length, the quantum potential term
$\tilde{\Box}\sqrt{\rho}/\sqrt{\rho}$ will be very small compared
with the classical kinetic term
$\tilde{g}^{ab}\partial_aS\partial_bS.$ We therefore can be neglect
it in the WKB approximation. In other words in the de Broglie-Bohm
particle interpretation the quantum potential causes to trajectories
on the classical paths of the particles ensemble. These particles
move normal to the wave front $S=constant.$ It follows then that the
Hamilton-Jacobi equation (2.22) can be regarded as a conservation
equation for the probability in an ensemble of such particles, all
moving normal to the same wave front with a probability density
$\rho.$ One should be note that the metric function $\tilde{g}_{ab}$
is still treats as classical metric field and whose quantum
trajectories come from quantum potential term
$\tilde{\Box}\sqrt{\rho}/\sqrt{\rho}$ such as follows. \\The
Hamilton Jacobi equation (2.22) can be rewritten as \be
\mathbf{g}^{ab}\partial_aS\partial_bS=-\alpha \tilde{R},\ee where we
defined quantum perturbed metric function as \be
\mathbf{g}_{ab}=\tilde{g}_{ab}\left\{1-\frac{\tilde{\Box}\sqrt{\rho}/\sqrt{\rho}}{\alpha
\tilde{R}}\right\}.\ee Thus quantum trajectories of the classical
metric will be \be
\Delta_{ab}=\mathbf{g}_{ab}-\tilde{g}_{ab}=-\tilde{g}_{ab}\left[
\frac{\tilde{\Box}\sqrt{\rho}/\sqrt{\rho}}{\alpha
\tilde{R}}\right]\ee which are negligible in the WKB approximation.
Interaction of the particles ensemble with the classical background
metric $\tilde{g}_{ab}$ causes to create these quantum trajectories.
We should be point  that the covariant conservation condition (2.23)
described in terms of the quantum perturbed metric function
$\mathbf{g}_{ab}$ will be violated. This violation originates from
creation of quantum particles interacting with curved space time
(Hawking radiation). Physically it seems that the pointed creation
of particles should be related to the conformal anomaly derived from
renormalization of stress tensor operator expectation value of the
quantum fields propagating on a curved space time (see [29] and
references therein). As a future work this problem may to be a
suitable exercise leading to a duality picture between de
Broglie-Bohm anomaly derived from quantum potential and conformal
anomaly obtained from renormalization theory of quantum fields.
\section{Concluding remarks}
Applying JBD gravity theory described by spherically symmetric
curved space time we seek corresponding  de Broglie Bohm pilot wave
perspective, where amplitude and phase part of the pilot wave is
defined by the Brans Dicke scalar field and 2-sphere scale factor.
This approach of quantum gravity is useful to study spherically
symmetric space time with black hole topology. Also quantum
trajectory of the background metric is obtained in terms of the Bohm
quantum potential. There is obtained a suitable conformal frame
where the covariant conservation of the matter-energy current
density of particles ensemble is established. However it violates in
the presence of the quantum potential effects.
 \vskip 0.1 cm {\bf References}
\begin{description}
\item[1.] B. Fauser, J. Tolksdorf and E. Zeidler, \textit{Quantum
Gravity-Mathematical Models And Experimental Bounds}, Birkh\"{a}user
verlag, Basel-Boston-Berlin (2007).
\item[2.] N. D. Birrell and P. C. W. Davies, \textit{Quantum
Fields In Curved Space }, Cambridge University press (1984).
\item[3.] L. Parker and D. Toms, \textit{Quantum Field Theory In Curved Space Time-Quantized Fields And Gravity}
, Cambridge University press (2009).
\item[4.] F. Bastianelli and P. Van Nieuwenhuizen, \textit{Path Integrals And Anomalies In Curved
Space}, Cambridge University press (2006).
\item[5.] B. R. Iyer, N. Mukunda and C. V. Vishveshwara, \textit{Gravitation, Gauge Theories And The Early
Universe}, Kluwer academic publishers (1989).
\item[6.] B. S. De Witt, Phys. Rev. D160, 1113, (1967).
\item [7.] T. Thiemann, \textit{Modern Canonical Quantum General
Relativity}, Cambridge University press (2008).
\item[8.] D. Bohm and B. J. Hiley, \textit{The Undivided Universe-An Ontological Interpretation Of Quantum Theory},
 Routledge, 11 New Fetter Lane, London EC4P, 4EE (1993).
\item[9.] P. R. Holland, \textit{The Quantum Theory of Motion-An Account Of The de Broglie Bohm Causal Interpretation Of Quantum Mechanics},
 Cambridge University press (1993).
 \item[10.] L. de Broglie, \textit{Non Linear Wave Mechanics,}
 translated to English, by A. J. Kondel, Elsevier publishing company
 (1960).
 \item[11.] M. Bell, K. Gottfried and M. Veltman, \textit{John S. Bell On The Foundations Of Quantum
 Mechanics}, World Scientific publishing Co. Pte. Ltd (2001).
\item[12.] G. Greenstein and A. G. Zajonc \textit{The Quantum
Challenge-Modern Research On The Foundations Of Quantum Mechanics},
Jones and Bartlett publishers, Inc (2006).
\item[13.] R. Vasudevan, K. V. Parthasarathy and R. Ramanathan, \textit{Quantum Mechanics-A Stochastic
Approach}, Alpha science international LTD, 7200. The Quorum, Oxford
Business Park North, Garsington Road, Oxford OX4 2JZ, U.K. (2008).
\item[14.] R. E. Wyatt, \textit{Quantum Dynamics With Trajectories-Introduction To Quantum
Hydrodynamics}, Springer Science+Business Media, Inc (2005).
\item[15.] J. Soda, H. Ishihara and O. Iguchi, arXiv: gr-qc/9509008 v1, 6 Sep. (1995).
\item[16.] M. Kenmoku, H. Kubotani, E. Takasugi and Y. Yamazaki,
arXiv: gr-qc/ 9810039 v1, 10 Oct (1998).
\item[17.] A. Blaut and J. K. Glikman, Class.Quantum Grav.13, 39
(1996).
\item[18.] El-Nabulsi Ahmad Rami, Rom. Journ. Phys. Vol.53, Nos. 7-8, P. 933-940,
Bucharest, (2008).
\item[19.] C. Brans and R. Dicke, Phys. Rev. 124, 925 (1961).
\item[20.] H. Ghafarnejad, Astrophys. Space Sci. 301, 145-148,
(2006).
\item[21.] Y. M. Cho, Class. Quantum. Grav.10, 2963 (1997).
\item[22.] C. M. Will, \textit{Theory And Experiment In Gravitational
Physics}, Cambridge University press (1993); revised version:
gr-qc/9811036.
\item[23.] E. Gaztanaga and J. A. Lobo, Astrophys. J., 548, 47
(2001).
\item[24.] R. D. Reasenberg et al, Astrophys. J., 234, 925 (1961).
\item[25.] C. M. Will, Living Rev. Rel. 9 (2006);
http://WWW.livingreviews.org/lrr-2006-3.
\item[26.] L. J. Garay and J. G. Bellido, gr-qc/9209015 (1992).
\item[27.] J. M. Overdid and P. S. Wesson, gr-qc/9805018 (1998).
\item[28.] P. Thomi, B. Isaac and P. Hajicek, Phys. Rev. D30, 1168
(1984).
\item[29.] H. Ghafarnerjad and H. Salehi, Phys. Rev. D56, 4633
(1997); Phys. Rev. D57, 5311 (1998).
  \end{description}
\end{document}